\documentclass[12pt]{article}

\usepackage{latexsym}
\usepackage{tikz}
\usetikzlibrary{arrows,automata}
\usepackage{graphicx}
\usepackage{dcolumn}

\usepackage{amssymb}
\usepackage{amsmath}

\newcommand{\bq}{\begin{equation}}
\newcommand{\eq}{\end{equation}}

\newcommand{\bqr}{\begin{eqnarray}}
\newcommand{\eqr}{\end{eqnarray}}

\newcommand{\bqrx}{\begin{eqnarray*}}
\newcommand{\eqrx}{\end{eqnarray*}}

\newcommand{\br}{\begin{array}}
\newcommand{\er}{\end{array}}

\newcommand{\blsk}{\baselineskip}

\usepackage[normalem]{ulem}

\begin{document}

\pagestyle{plain}
\pagenumbering{arabic}

\setlength{\parindent}{0in}
\setlength{\parskip}{3ex}
\setlength{\footskip}{.6in}



\setlength{\blsk}{.25in}%
\vspace*{.6in}
\begin{center}
Probability distributions for Markov chains based quantum walks
\end{center}
\begin{center}

Radhakrishnan\, Balu, Chaobin\, Liu, Salvador\, E. Venegas-Andraca
  \footnote{radhakrishnan.balu.civ@mail.mil, U.S. Army Research Laboratory, Computational and Information Sciences Directorate, Adelphi, MD 20783, USA\\
radbalu1@umbc.edu, Computer Science and Electrical Engineering, University
of Maryland Baltimore County, 1000 Hilltop Circle, Baltimore, MD
21250, USA\\

cliu@bowiestate.edu, Department of Mathematics, Bowie State University, Bowie, MD 20715, USA\\

salvador.venegas-andraca@keble.oxon.org, Tecnologico de Monterrey, Escuela de Ingenieria y Ciencias, Ave. Eugenio Garza Sada 2501, Monterrey, NL 64849, Mexico}
\end{center}

\begin{abstract}
{We analyze the probability distributions of the quantum walks induced from Markov chains by Szegedy (2004). The first part of this paper is devoted to the quantum walks induced from finite state Markov chains. It is shown that the probability distribution on the states of the underlying Markov chain is always convergent in the Cesaro sense. In particular, we deduce that the limiting distribution is uniform if the transition matrix is symmetric. In the cases of non-symmetric Markov chain, we exemplify that the limiting distribution of the quantum walk is not necessarily identical with the stationary distribution of the underlying irreducible Markov chain. The Szegedy scheme can be extended to infinite state Markov chains (random walks). In the second part, we formulate the quantum walk induced from a lazy random walk on the line. We then obtain the weak limit of the quantum walk. It is noted that the current quantum walk appears to spread faster than its counterpart-quantum walk on the line driven by the Grover coin discussed in literature. The paper closes with an outlook on possible future directions.
 }
\end{abstract}%

keywords: Szegedy quantum walks, Markov chains, asymptotic distributions, weak limits.

\section{Introduction}

Random walks have proved to be a fundamental mathematical tool for modeling and simulating complex problems and natural phenomena. Among the various applications of random walks we find the development of stochastic algorithms \cite{motwani95,mitzenmacher05} for problems of paramount importance in theoretical computer science \cite{schoning99}, earthquake modeling \cite{minson13}, computer vision \cite{grady06}, financial modeling \cite{fama95}, graph theory \cite{pons06}, proteomics \cite{dotu11,madras96} and Internet technology \cite{brin98a,brin98b,brin99,bianchini05}.


As a quantum mechanical counterpart of random walks, in recent times quantum walks \cite{ADZ93, M96, FG98, AAKV01, ABNVW01, W2001,S04} have been extensively studied. Quantum walks were originally devised under the same rationale as classical random walks, as a mathematical basis to develop sophisticated algorithms (e.g. \cite{ CCDF02,     
SKW03,ambainis04a, AA07,ambainis08_sofsem,mohseni08}). Unlike the random walks formulated by iteration of probability transition matrices on a probability distribution, quantum walks are defined by unitary evolutions of the {\em probability amplitudes}. The resulting probability distribution is defined to be the sum of squares of the norms of amplitudes so that there exists a non-linearity map between the initial state and the resulting probability distribution. Due to quantum interference effects, quantum walks have been shown to outperform random walks at certain computational tasks \cite{CCDF02, SKW03, AA07, CG04a, CSV07, AKR05, CG04b, T08, FGG08, YPSP15}; moreover, it has been proved that both continuous and discrete quantum walks constitute universal models of quantum computation \cite{AMC2009,LCETK10}. For a lively and informative elaboration of the history of quantum walks and their connection to quantum computing, physics and the natural sciences, the reader is referred to \cite{K03,K06,K08,VA08, MB11, VA12,portugal13} and the references therein.

We distinguish between two types of QW: {\em continuous time} and {\em discrete time}. In both cases, quantum walks are run on graphs (i.e., on discrete spaces) but their evolution timing is different: as for continuous quantum walks, evolution is described by the Schr\"odinger equation whose Hamiltonian is built based on the Laplacian matrix of the graph where the quantum walk is run on \cite{FG98}; as for discrete quantum walks, evolution is described by unitary operators that are applied in discrete time steps \cite{AAKV01,ABNVW01, S04}. 


As for discrete-time quantum walks, in addition to the ``position" register, an extra register is needed to store the direction in which the walker unitarily moves from a node to its neighboring nodes. Two important types of discrete-time quantum walks have been studied:

\begin{itemize}

\item {\bf Coin-driven quantum walks (coin-driven QW)}. This model of quantum walk, initially proposed by \cite{AAKV01, ABNVW01}, is composed of two quantum systems:  i) a walker, which is a quantum system living in a Hilbert space of finite or infinite but countable dimension ${\cal H}_p$, and ii) a coin, which is a quantum system living in a 2-dimensional Hilbert space ${\cal H}_c$. The unitary maps in the framework of coin-driven QW are defined in terms of conditional shifts and coin operators, by which the evolution of the pure state of the quantum system is governed. 


\item {\bf Markov chain-based quantum walks (MCBQW)}. The other type of discrete-time quantum walk was proposed by \cite{S04, AA07} where both registers are nodes. The corresponding unitary map in this scenario is given by a swap operator and a {\em reflection operator} in the Hilbert space. It is noted that the reflection operator is derived by {\em quantizing} a Markov chain associated with the underlying graph. 

\end{itemize}


Although MCBQW have been studied in some detail by the scientific community (see, for example,  \cite{MNPS11, KMOR10, SP10,BS06, LW2017} and the references therein), many mathematical, statistical and computational properties  of MCBQW remain to be discovered, like the asymptotic average probability distributions of MCBQW on a general finite graph that, so far, remain largely unexplored. Moreover, results on probability distributions and properties of the underlying Markov chains have been barely explored. For instance:

\begin{itemize}

 \item It is unknown whether MCBQW constitute a model of universal quantum computation or if it is equivalent to a more modest model of quantum computation like quantum annealing \cite{kadowaki98}.

\item
In \cite{PM12}, the authors propose a class of quantum PageRank algorithms and introduce an instance of this class. Since the original formulation of PageRank can be described in terms of Markov chains \cite{brin98a,brin98b,brin99,bianchini05}, the class of quantum PageRank algorithms introduced in \cite{PM12} is required to admit a quantized Markov chain description. For the purpose of mathematically characterizing the behavior of this quantum PageRank instance, two notions are proposed: the instantaneous importance $I_q(P_i,m)$  of a quantum web page $P_i$ (that is, the quantum PageRank of $P_i$) be equal to the probability of finding the quantum walker at node $i$  after $m$ time steps. Due to unitary evolution, $I_q(P_i,m)$ does not converge,  hence the notion of average probability $ \bar{P}_T(j|\alpha_0)$ is introduced.  Interesting numerical results related to the properties of quantum PageRank on some graphs are presented; however, no asymptotic analytical results are shown. 


\item
In \cite{IKS12}, although the asymptotic average probability distributions were derived, their discussion is only confined to the MCBQW on a finite {\em path}.

\item 
In \cite{S13}, the author managed to treat the relationship between the localization of MCBQW and the recurrent properties of the underlying Markov chains on a {\em half line}. To the best of our knowledge, this is the only publication in which the inherent connection between the statistical properties of MCBQW and the properties of underlying Markov chains are examined.

\end{itemize}

As an effort to address important fronts related to MCBQW, in this paper we analyze the asymptotic probability distributions of MCBQW. In particular, we deduce that the limiting distribution is uniform if the transition matrix is symmetric. In the cases of non-symmetric Markov chain, we exemplify that the limiting distribution of the quantum walk is not necessarily identical with the stationary distribution of the underlying irreducible Markov chain. The Szegedy scheme \cite{S04} can be extended to infinite state Markov chains. In the second part, we formulate the quantum walk induced from a lazy random walk on the line. We then obtain the weak limit of the quantum walk. It is noted that the quantum walk appears to spread faster than its counterpart-quantum walk on the line driven by the Grover coin discussed in literature. The paper closes with an outlook on possible future directions.
 


\section{Preliminaries}


Szegedy developed a general method for quantizing a Markov chain to create a discrete-time quantum walk \cite{S04}. Let $P=(p_{jk})$ be an $n\times n$ stochastic matrix representing the transition probability matrix of a Markov chain on a directed graph $G(V, E)$. Here $V$ is the set of the vertices of $G$, $E$ is the set of the oriented edge of $G$, $E=\{(j,k): j,k \in V\}$. In order to introduce a discrete-time quantum walk based on the aforesaid Markov chain, we use as the Hilbert space of the quantum walk the span of all vectors representing the $n\times n$ (directed) edges of the graphs, i.e., $\mathcal{H}=\mathrm{span}\{|j\rangle\otimes |k\rangle, j,k \in V\}$. For simplicity, from now on the vectors (states) in $\mathcal{H}$ of the form $|j\rangle\otimes |k\rangle$ will be abbreviated to $|jk\rangle$. Sometimes inserting a $\otimes$ will be used to make things clearer.

Let us define the vector states: $|\psi_j\rangle=\sum_{k=1}^{n}\sqrt{p_{jk}}|jk\rangle$, which is a superposition of the vectors representing the edges outgoing from the $j^{th}$ vertex. The weights are given by (the square root of the entries of) the transition matrix $P$.

One can easily verify that due to the stochasticity of $P$, $\{|\psi_j\rangle\}_{j=1}^{n}$ is an orthonormal set in $\mathcal{H}$. The operator $\Pi=\sum_{j=1}^n|\psi_j\rangle\langle \psi_j|$, is then an orthogonal projector onto the subspace $\mathcal{H}_{\psi}=\mathrm{span}\{|\psi_j\rangle: j\in V\}$.  With this, a single step of the quantum walk is then given by the unitary operator $U:=S(2\Pi-1)$ where $S=\sum_{j,k}|jk\rangle\langle kj|$ is the swap operator. 

Definition 1.\, Given $|\alpha_0\rangle \in \mathcal{H}$, where $\||\alpha_0\rangle\|=1$, the expression $|\alpha_t\rangle=U^t|\alpha_0\rangle$ is called the state for the walk at time $t$. The corresponding quantum walk (MCBQW) with the initial state $|\alpha_0\rangle$ is represented by the sequence $\{|\alpha_t\rangle\}_{t=0}^{\infty}$. 

Indeed, the structure of operator $U:=S(2\Pi-1)$ resembles Grover's diffusion operator \cite{grover96}. Analysis on this regard have been presented in literature such as \cite{santos16,wong17a}.

As shown above, MCBQW (also called Szegedy quantum walk or Szegedy walks in literature) is based on the Markov chain on a directed graph, but it is not exactly defined on the graph. MCBQW may be thought of as a walk on the edges of the original graph, rather than on its vertices \cite{MS08}. It is worth to note that some publications call the physical topology of MCBQW {\it bipartite graph} (or {\it bipartite double cover}) of the original graph, see, for example \cite{santos16}, a step-by-step introduction to Szegedy's quantum walks (including the bipartite graph) together with a detailed example of this kind of quantum walks on a $3 \times 3$ lattice is presented in \cite{wong17b}. In what follows, without causing confusion, we may sometimes use the phrase ``MCBQW on a graph" just for simplicity's sake.


In order to analyze the MCBQW, we need to study the spectral properties of an  $n\times n$ matrix $D=(d_{jk})$, which can be viewed as a linear transformation on the space $\mathcal{H}_V=\mathbb{C}^n=\mathrm{span}\{|j\rangle: j\in V\}$ and indeed builds a bridge from the classical Markov chain to the quantum walk. This matrix is defined as follows:

\begin{equation}
d_{jk}=\sqrt{p_{jk}p_{kj}}.\label{matrixD}
\end{equation}

Let us then define an operator $A$ from the space $\mathcal{H}_V$ to $\mathcal{H}_{\psi}$:

$$A=\sum_{j=1}^{n}|\psi_j\rangle\langle j|$$

The following identities describe the relationships among these operators:

$$A^{\dagger}A=\mathbb{I}, AA^{\dagger}=\Pi, A^{\dagger}SA=D$$

Since $D$ is symmetric by its definition, without loss of the generality, we may assume that, via the Spectral Decomposition, $D=\sum_r \lambda_r|w_r\rangle \langle w_r|+\sum_{s}|u_s\rangle \langle u_s|-\sum_t|v_t\rangle \langle v_t\rangle$ where $\lambda_r (\ne \pm 1)$, $1$ and $-1$ are the eigenvalues of $D$, $\{|w_r\rangle, |u_s\rangle, |v_t\rangle\} $ is an {\it orthonormal} basis for $\mathcal{H}_V$. Each $|w_r\rangle$ is an eigenvector of $D$ with eigenvalue $\lambda_r$, $|u_s\rangle$ and $|v_t\rangle$ are eigenvectors of $D$ with eigenvalues $1$ and $-1$, respectively. It should be pointed out that each $\lambda_r\in (-1,1)$ by the construction of the matrix $D$. Since $A^{\dagger}A=\mathbb{I}$, $\{A|w_r\rangle, A|u_s\rangle, A|v_t\rangle\}$ is an orthonormal basis for $\mathcal{H}_{\psi}$. Thus, the subspace $\mathcal{H}_{\psi, S}=\mathrm{span}\{|\psi_j\rangle, S|\psi_j\rangle:j\in V\}$ is identical with the subspace $\mathrm{span}\{A|w_r\rangle, SA|w_r\rangle, A|u_s\rangle, SA|u_s\rangle, A|v_t\rangle, SA|v_t\rangle\}$, which is invariant under the unitary operator $U$. The spectral structure of unitary operator $U$ is intimately connected with the matrix $D$:

\begin{enumerate}
\item  $A|w\rangle-e^{\pm i \arccos \lambda_r}SA|w\rangle$ are eigenvectors of $U$ with eigenvalues $e^{\pm i \arccos \lambda_r}$, respectively.
\item  $A|u\rangle=SA|u\rangle$, and $A|u\rangle$ is an eigenvector of $U$ with eigenvalue $1$.
\item $A|v\rangle=-SA|v\rangle$, and  $A|v\rangle $ is an eigenvector of $U$ with eigenvalue $-1$.
\end{enumerate}

Items 2 and 3 imply that the invariant subspace of $U$,
$\mathcal{H}_{\psi, S}=\mathrm{span}\{A|w_r\rangle, SA|w_r\rangle, A|u_s\rangle, A|v_t\rangle\}$.

 It is straightforward to verify that $\{A|w_r\rangle-e^{\pm i \arccos \lambda_r}SA|w_r\rangle, A|u_s\rangle, A|v_t\rangle\}$, the collection of all these eigenvectors of $U$, forms an orthogonal set. Moreover, $\|A|u\rangle\|=\|A|v\rangle\|=1$ and $\|A|w_r\rangle-e^{\pm i \arccos \lambda_r}SA|w_r\rangle\|=\sqrt{2-2\lambda_r^2}$.

Since the set of vectors $A|w_r\rangle, SA|w_r\rangle, A|u_s\rangle, A|v_t\rangle$ are linearly independent, $\mathcal{H}_{\psi, S}$ is in fact identical with the subspace

$\mathrm{span}\{A|w_r\rangle-e^{\pm i \arccos \lambda_r}SA|w_r\rangle, A|u_s\rangle, A|v_t\rangle\}$, 

i.e., $\mathcal{H}_{\psi, S}=\mathrm{span}\{A|w_r\rangle-e^{\pm i \arccos \lambda_r}SA|w_r\rangle, A|u_s\rangle, A|v_t\rangle\}$.



We set 

\begin{enumerate}
  \item  $A|w_r^+\rangle=(A|w_r\rangle-e^{i \arccos \lambda_r}SA|w_r\rangle/\sqrt{2-2\lambda_r^2}$
  \item  $A|w_r^-\rangle=(A|w_r\rangle-e^{-i \arccos \lambda_r}SA|w_r\rangle)/\sqrt{2-2\lambda_r^2}$
\end{enumerate}


Then the invariant subspace of $U$ is identical to the subspace $\mathrm{span}\{A|w_r^+\rangle, A|w_r^-\rangle, A|u_s\rangle, A|v_t\rangle\}$. This is a subspace spanned by the set of orthonormal eigenvectors of $U$ associated with the key operator $D$.

Let us decompose the Hilbert space $\mathcal{H}$ into $\mathcal{H}_{\psi,S}$ and its orthogonal complement $\mathcal{H}_{\psi,S}^{\bot}$, i.e., $\mathcal{H}=\mathcal{H}_{\psi,S}\oplus\mathcal{H}_{\psi, S}^{\bot}$. It is not difficult to check that the action of $U$ on $\mathcal{H}_{\psi,S}^{\bot}$ is exactly $-S$ and this subspace is invariant under $U$, thereby $U^2$ just trivially acts on the subspace as an identity. The nontrivial dynamics of $U$ only takes place on the subspace $\mathcal{H}_{\psi,S}$. By its construction, the dimension of the subspace  $\mathcal{H}_{\psi,S}$ is at most $2n$ (the dimension of the whole space $\mathcal{H}$ is $n^2$), which can be achieved only if $D$ does not have both $1$ and $-1$ as its eigenvalues. Based on the aforesaid observation, we may confine the initial state of the quantum walk to the subspace $\mathcal{H}_{\psi, S}$, which is spanned by the set of the orthonormal eigenvectors of $U$: $\{A|w_r^+\rangle, A|w_r^-\rangle, A|u_s\rangle, A|v_t\rangle\}$.

For the sake of better exposure of our main result, we relabel the above orthonormal eigenvectors of $U$, which forms a basis for the invariant subspace $\mathcal{H}_{\psi,S}$, as $\{|\phi_l\rangle\}$ with associated eigenvalues $\{\mu_l\}, l=1,2,..., m$ where $m\le 2n$. 



\section{Asymptotic distribution on the states of the underlying Markov chain}

We now proceed to study the evolution of the quantum walk as a function of time. Provided that the initial state $|\alpha_0\rangle=\frac{1}{\sqrt{n}}\sum_{j=1}^n|\psi_j\rangle$, then the state of the quantum walk at time $t$ is $|\alpha_t\rangle=U^{t}|\alpha_0\rangle$. Since $U$ is unitary, in general $\lim_{t \rightarrow \infty} |\alpha_t\rangle$ does not exist. Now consider instead the probability distribution on the vertices $\{|j\rangle:j\in V\}$ of the underlying graph induced by states of the quantum walks $|\alpha_t\rangle$, and let $P_t(j|\alpha_0)$ denote the probability of finding the walker at vertex $|j\rangle$ at time $t$.

Definition 2.\, $P_t(j|\alpha_0)=\sum_{k}|\langle jk|\alpha_t\rangle|^2$.

As a matter of fact, $P_t$ does not converge either. However, the average of $P_t$ over time would converge as Theorem 1 below shows. The limit of the average of $P_t$ is called the asymptotic average probability distribution. We define:

Definition 3.\, $\bar{P}_T(j|\alpha_0)=\frac{1}{T}\sum_{t=1}^T P_t(j|\alpha_0)$, $\bar{P}_{\infty}(j|\alpha_0)=\lim_{T\mapsto \infty}\bar{P}_T(j|\alpha_0)$.

We shall give an explicit formula for the limit of $\bar{P}_{\infty}$.

Theorem 1\, Given a Markov chain on the state space $V$ with the transition matrix $P$, the induced quantum walk is defined as $|\alpha_t\rangle=U^{t}|\alpha_0\rangle$ where the initial state $|\alpha_0\rangle=\sum_l\langle \phi_l|\alpha_0\rangle |\phi_l\rangle$, then 
\begin{eqnarray}
\bar{P}_{\infty}(j|\alpha_0)=\lim_{T\mapsto \infty}\bar{P}_T(j|\alpha_0)=\sum_k\sum_{l, m}\langle \phi_l|\alpha_0\rangle \langle jk|\phi_l\rangle\langle \alpha_0| \phi_m\rangle \langle \phi_m |jk\rangle \label{MCBQW_prob}
\end{eqnarray}
where the first sum is over all values of $k$, and the second sum is only on pairs $l, m$ such that $\mu_l=\mu_m$.

Proof\,\, Note that $|\alpha_t\rangle=U^t|\alpha_0\rangle=\sum_l\langle\phi_l|\alpha_0\rangle\mu_l^t|\phi_l\rangle$, then one has 
$$|\alpha_t\rangle\langle\alpha_t|=\sum_l\sum_m\langle\phi_l|\alpha_0\rangle\overline{\langle\phi_m|\alpha_0\rangle}(\mu_l\bar{\mu}_m)^t|\phi_l\rangle\langle\phi_m|.$$
Since $\bar{P}_T(j|\alpha_0)=\frac{1}{T}\sum_{t=1}^T P_t(j|\alpha_0)=\frac{1}{T}\sum_{t=1}^T\sum_{k}|\langle jk|\alpha_t\rangle|^2$, 
\begin{equation}
\bar{P}_T(j|\alpha_0)=\frac{1}{T}\sum_{t=1}^T\sum_{k}\sum_l\sum_m\langle\phi_l|\alpha_0\rangle\overline{\langle\phi_m|\alpha_0\rangle}(\mu_l\bar{\mu}_m)^t\langle jk|\phi_l\rangle\langle\phi_m|jk\rangle \label{meanprob}
\end{equation}
We separate the sum in the right-hand side of Eq.(\ref{meanprob}) into two parts: One part in which $\mu_l=\mu_m$, this part is $\sum_{k}\sum_l\sum_m\langle\phi_l|\alpha_0\rangle\overline{\langle\phi_m|\alpha_0\rangle}\langle jk|\phi_l\rangle\langle\phi_m|jk\rangle$. The other part in which $\mu_l\ne\mu_m$ can be written as $\sum_{k}\sum_l\sum_m\frac{\mu_l\bar{\mu}_m[1-(\mu_l\bar{\mu}_m)^T]}{T(1-\mu_l\bar{\mu}_m)}\langle\phi_l|\alpha_0\rangle\overline{\langle\phi_m|\alpha_0\rangle}\langle jk|\phi_l\rangle\langle\phi_m|jk\rangle$.

It can be seen that the latter part converges to zero as $T$ goes to infinity. The only contribution to the limit of the average probability $\bar{P}_T(j|\alpha_0)$ comes from the part with $\mu_l=\mu_m$. This completes the proof.  

When the transition matrix $P$ of a Markov chain is symmetric, the limiting probability is uniform. This assertion is recorded in the theorem below.

Theorem 2\,\, Given a Markov chain on the state space $V$ with a symmetric transition matrix $P$, the induced quantum walk is defined as $|\alpha_t\rangle=U^{t}|\alpha_0\rangle$ where the initial state $|\alpha_0\rangle=\sum_l\langle \phi_l|\alpha_0\rangle |\phi_l\rangle$, then 
$\lim_{T\mapsto \infty}\bar{P}_T(j|\alpha_0)=\frac{1}{n}$ where $n$ is the size of the transition matrix.

To prove theorem 2, we need the following three facts about the matrix $D$ and a symmetric transition matrix $P$:

Lemma 1\,\, Provided that $w$ ($=[w(1), w(2),..., w(n)]^T$) is an eigenvector of $D$ with the corresponding eigenvalue $\lambda\ne \pm 1$, then it is true that $\langle A^{\dagger}S\alpha_0, w\rangle=\lambda\frac{\sum_jw(j)}{\sqrt{n}}$ where $\langle X, Y\rangle$ is the dot product in the Euclidean space $\mathbb{R}^n$.

Proof\,\, Note that $\langle \psi_j|S\psi_l\rangle=\langle \sum_k\sqrt{p_{jk}}|jk\rangle, \sum_k\sqrt{p_{lk}}|kl\rangle\rangle=\sqrt{p_{jl}p_{lj}}$. Then we have 
$\langle A^{\dagger}S\alpha_0, w\rangle=\langle\sum_j|j\rangle\langle\psi_j|S\frac{1}{\sqrt{n}}\sum_l|\psi_l\rangle, w\rangle=\frac{1}{\sqrt{n}}\sum_j[\sum_ld_{jl}w(l)]=\lambda\frac{\sum_jw(j)}{\sqrt{n}}$

Lemma 2\,\, If $P$ is symmetric and $w$ is an eigenvector corresponding to the eigenvalue $\lambda\ne 1$, then $\sum_j w(j)=0$.

Proof\,\, We denote $\vec{1}=(1, 1, ..., 1)^{\mathrm{T}}$. Note that $\sum_j w(j)={\vec{1}}^{\mathrm{T}}Pw=\lambda\vec{1}w=\lambda\sum_jw(j)$, this implies that $\sum_jw(j)=0$.

Lemma 3\,\, Provided that $P$ is symmetric. $u_0=\frac{1}{\sqrt{n}}\vec{1}$. If $u$ is also an eigenvector of $P$ with eigenvalue $1$ such that the dot product $\langle u, u_0\rangle=0$, then $\langle \alpha_0|Au\rangle=0$.

Proof\,\, Note that $\sum_j u(j)=0$ and $\langle \alpha_0|Au\rangle=\sum_j u(j)/\sqrt{n}$, so $\langle \alpha_0|Au \rangle=0$.

Proof of theorem 2\,\,\, To make the proof more readable, we recall some notations used before. For the given transition matrix $P$, the associated matrix $D$ is defined by Eq.(\ref{matrixD}). Its spectral decomposition is assumed to be $D=\sum_r \lambda_r|w_r\rangle \langle w_r|+\sum_{s}|u_s\rangle \langle u_s|-\sum_t|v_t\rangle \langle v_t\rangle$ where $\lambda_r (\ne \pm 1)$, $1$ and $-1$ are the eigenvalues of $D$, $\{|w_r\rangle, |u_s\rangle, |v_t\rangle\} $ is an {\it orthonormal} basis for $\mathcal{H}_V$. Each $|w_r\rangle$ is an eigenvector of $D$ with eigenvalue $\lambda_r$, $|u_s\rangle$ and $|v_t\rangle$ are eigenvectors of $D$ with eigenvalues $1$ and $-1$, respectively. 







We compute the values of $\langle \alpha_0|\phi\rangle$ and $\langle jk|\phi\rangle$ as follows:

1)\,$|\phi\rangle$ is an eigenvector of $U$ with the corresponding eigenvalue 1. In this case, $|\phi\rangle=A|u\rangle$ where $|u\rangle$ is an eigenvector of $D$ with the corresponding eigenvalue 1. Then $\langle \alpha_0|\phi\rangle=\langle \frac{1}{\sqrt{n}}\sum_{j=1}^n|\psi_j\rangle|A|u\rangle=\sum_j u(j)/\sqrt{n}$.

2)\, $|\phi\rangle$ is an eigenvector of $U$ with the corresponding eigenvalue $-1$. In this case, $|\phi\rangle=A|v\rangle$ where $|v\rangle$ is an eigenvector of $D$ with the corresponding eigenvalue $-1$. Then $\langle \alpha_0|\phi\rangle=\langle \frac{1}{\sqrt{n}}\sum_{j=1}^n|\psi_j\rangle|A|v\rangle=\sum_j v(j)/\sqrt{n}$.

3)\, $|\phi\rangle$ is an eigenvector of $U$ with the corresponding eigenvalue $e^{\pm i \arccos \lambda_r}(\ne \pm1)$. In this case, $|\phi\rangle=A|w\rangle-e^{\pm i \arccos \lambda}SA|w\rangle$ where $|w\rangle$ is an eigenvector of $D$ with the corresponding eigenvalue $\lambda$. Then, by Lemma 1, we have $\langle \alpha_0|\phi\rangle=\langle \frac{1}{\sqrt{n}}\sum_{j=1}^n|\psi_j\rangle|A|w\rangle-e^{\pm i \arccos \lambda_r}SA|w\rangle=\frac{\sum_j w(j)(1-\lambda e^{-i\arccos(\lambda)})}{\sqrt{2(1-\lambda^2)n}}$.

4)\, $|\phi\rangle$ is an eigenvector of $U$ with the corresponding eigenvalue 1. In this case, $|\phi\rangle=A|u\rangle$ where $|u\rangle$ is an eigenvector of $D$ with the corresponding eigenvalue 1. Then $\langle jk|\phi\rangle=\langle A^{\dagger}jk|u\rangle=\sqrt{p_{jk}}u(j)$.

Applying the values of $\langle \alpha_0|\phi\rangle$ we computed above and Lemma 2, we conclude that $$\lim_{T\mapsto \infty}\bar{P}_T(j|\alpha_0)=\sum_k\sum_{l, m}\langle \alpha_0|\phi_l\rangle \langle jk|\phi_l\rangle\langle \phi_m| \alpha_0\rangle \langle \phi_m |jk\rangle$$ where the first sum is over all values of $k$, and the second sum is only on pairs $l, m$ such that $\mu_l=\mu_m=1$.

By Lemma 3, the aforesaid sum is further reduced to be 
$$\bar{P}_{\infty}(j|\alpha_0)=\sum_k\langle \alpha_0|Au_0\rangle \langle jk|Au_0\rangle\langle Au_0| \alpha_0\rangle \langle Au_0 |jk\rangle$$
where the sum is over all values of $k$, and $u_0=\frac{1}{\sqrt{n}}(1,1,...,1)^{\mathrm{T}}$. Applying the items 1 and 4 to $\bar{P}_{\infty}(j|\alpha_0)$, we have that $\bar{P}_{\infty}(j|\alpha_0)=\frac{1}{n}$. This completes the proof.


Results presented on Theorems 1 and 2 encourage us to pursue further analytical expressions for long-term behavior of probability distributions as, in addition to mathematically characterizing quantum Markov chains on different graphs, those expressions will be useful for describing the asymptotic behavior of new quantum Markov chain-based algorithms. As an example of further developments, we envision the analysis of non-regular graphs for quantum PageRank algorithms like those numerically studied in \cite{PM12,PMCM13}). 

Having analyzed and obtained the asymptotic average probability distributions (AAPD) of MCBQW, we might want to understand how the properties of AAPD can be reflected by the property of the underlying Markov chain, for instance, is the AAPD of MCBQW identical with the stationary distribution of the underlying Markov chains? To this end, we shall conduct a study on both MCBQW and Markov chains over some graphs.

To improve the readability of our discussion on the study, we gather the things related to the probability distributions and stationary distributions of both MCBQW and its underlying Markov chain as follows:


A Markov chain on a directed graph $G$ can be described by its transition probability matrix $P $ in which the entry $p_{jk}$ represents the probability of making a transition to vertex $k$ from vertex $j$. It should be pointed out that to preserve normalization, we must have $\sum_{k}p_{jk}=1$.  Let $u$ be the probability vector (horizontal) which represents the starting distribution at the vertices of the graph. Then the probability distribution after one step of the walk becomes $uP$. If $uP=u$, then $u$ is called a stationary distribution of the Markov chain on the graph, which is often denoted by $\pi$. In the following study, let us adopt the usual Markov chains associated with $G$ such that its transition matrix of the Markov chain is $P=D_{in}^{-1}A$,  where $A$ is the adjacency matrix associated with the edges incident at various vertices, and where $D_{in}$ is  the degree matrix associated with edges incident at various vertices. 

For the Markov chains on a directed graph $G$ given by the transition probability matrix $P$, MCBQW is defined by the sequence $\{|\alpha_t\rangle\}_{t=0}^{\infty}$ (for details, please refer to Definition 1). The probability of finding the quantum walker at vertex $j$ at time $t$ is given by $P_t(j|\alpha_0)=\sum_{k}|\langle jk|\alpha_t\rangle|^2$, and the asymptotic average probability of finding the quantum walker at the $|j\rangle$ is given by $\bar{P}_{\infty}(j|\alpha_0)=\lim_{T\mapsto \infty}\frac{1}{T}\sum_{t=1}^T P_t(j|\alpha_0)$ (for details, please refer to Definitions 2 and 3), which can be calculated by Eq.(\ref{MCBQW_prob}) in theorem 1. We have to admit that the calculation would be tedious.

To avoid unpleasant complications and to permit us more easily to illustrate the asymptotic average probability distribution of MCBQW and some basic attributes of the underlying Markov chains \cite{AF2014}, we concede, in this study, to confine our attention to the quantum walks and Markov chains on four simple directed graphs (see Figure \ref{fig:MC_{1234}}). A summary of the study are shown in the table below.



\begin{figure}
\begin{center}
 \begin{tikzpicture}[->,>=stealth',shorten >=1pt,auto,node distance=2.8cm,
                    semithick]
  \tikzstyle{every state}=[fill=red,draw=none,text=white]

  \node[state] (A)                    {$v_2$};
  \node[state]         (B) [left of=A] {$v_1$};
  \node[below] at (current bounding box.south) {Graph 1};

  \path (A) edge              node {} (B)
        (B) edge [loop above] node {} (B);
    \end{tikzpicture} \hspace{1cm} \begin{tikzpicture}[->,>=stealth',shorten >=1pt,auto,node distance=2.8cm,
                    semithick]
  \tikzstyle{every state}=[fill=red,draw=none,text=white]

  \node[state] (A)                    {$v_2$};
  \node[state]         (B) [left of=A] {$v_1$};
  \node[below] at (current bounding box.south) {Graph 2};

  \path (A) edge [bend right]             node[above] {} (B)
        (B) edge [bend right]             node[below] {} (A)
        (B) edge [loop above] node {} (B);
\end{tikzpicture}

\vskip 2cm

\begin{tikzpicture}[->,>=stealth',shorten >=1pt,auto,node distance=2.8cm,
                    semithick]
  \tikzstyle{every state}=[fill=red,draw=none,text=white]

  \node[state] (C) {$v_1$};
    \node[state] (A) [below right of=C]                   {$v_2$};
  \node[state]         (B) [below left of=C] {$v_3$};
  \node[below] at (current bounding box.south) {Graph 3};

  \path (A) edge  [bend right]                  node {}(C)
        (C) edge                                node[above] {} (B)
        (C) edge [bend right]             node[below] {} (A)
        (B) edge                          node {} (A);

\end{tikzpicture}\hspace{1cm}\begin{tikzpicture}[->,>=stealth',shorten >=1pt,auto,node distance=2.8cm,
                    semithick]
  \tikzstyle{every state}=[fill=red,draw=none,text=white]

  \node[state] (C) {$v_1$};
    \node[state] (A) [right of=C]                   {$v_2$};
  \node[state]         (B) [right of=A] {$v_3$};
  \node[below] at (current bounding box.south) {Graph 4};

  \path (A) edge  [bend right]                  node {}(C)
        (C) edge   [bend right]           node[below] {} (A)
         (A) edge   [bend right]          node {} (B)
        (B) edge   [bend right]                       node {} (A);

\end{tikzpicture}

\caption{Four Directed Graphs}  \label{fig:MC_{1234}}
\end{center}
\end{figure}
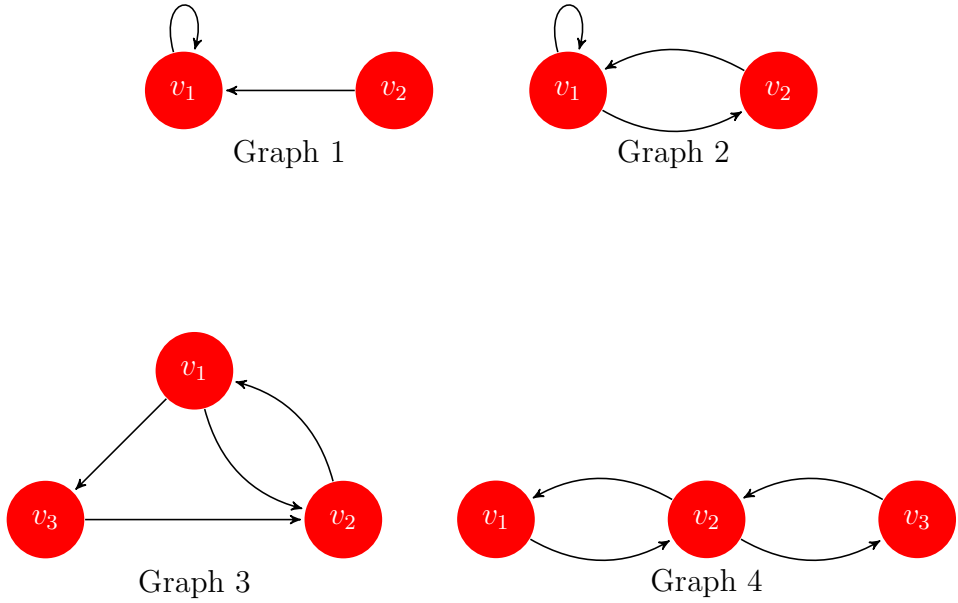






\begin{center}
{\bf Table I Probability Distributions: MCBQW vs MC} 
\begingroup
\renewcommand{\arraystretch}{4}

 \begin{tabular}{|*5{>{\renewcommand{\arraystretch}{1}}c|}}
\hline
&\text{Graph 1} & \text{Graph 2}&\text{Graph 3}&Graph 4\\
\hline
\textbf{$P$ (MC)}&$\left[ \begin{array}{cc} 1 & 0  \\ 1 & 0 \end{array}\right]$ & $\left[ \begin{array}{cc} .5 & .5 \\ 1 & 0  \end{array}\right]$&$\left[ \begin{array}{ccc} 0&.5 & .5 \\ 1 & 0&0\\0&1&0 \end{array}\right]$&$\left[ \begin{array}{ccc} 0& 1 & 0  \\ .5&0& .5\\0&1&0\end{array}\right]$\\
\hline
\textbf{Properties of MC.}&redu, reve & ergodic, reve&ergodic, not reve&irred, periodic, reve\\
\hline
\textbf{$\pi$ (MC)}&$(1,0)$ & $(\frac{2}{3},\frac{1}{3})$ &$(\frac{2}{5},\frac{2}{5},\frac{1}{5})$&$(\frac{1}{4},\frac{1}{2},\frac{1}{4})$\\
\hline
\textbf{$\bar{P}_{\infty}$ (QW})&$(\frac{3}{4},\frac{1}{4})$ & $ (\frac{2}{3},\frac{1}{3})$ &$(\frac{1}{3},\frac{1}{3},\frac{1}{3})$&$(\frac{1}{4},\frac{1}{2},\frac{1}{4})$\\
\hline
\end{tabular}
\endgroup
\end{center}

{\bf Notes}:\,\, In this table, $P$ is the transition matrix of the Markov chain associated with a graph, $\pi$ is the stationary probability distribution of a Markov chain, $\bar{P}_{\infty}$ is the asymptotic average probability distribution of MCBQW, ``redu" stands for ``reducible", ``irred" stands for ``irreducible", and ``reve" stands for ``reversible". 



\section{Weak limits for MCBQW on the line}

\begin{figure}
\begin{center}
\begin{tikzpicture}[->,>=stealth',shorten >=1pt,auto,node distance=1.8cm,
semithick]
\tikzstyle{every state}=[fill=red,draw=none,text=white, scale=1]
\node                      (l)                {$\cdots$};
\node[state]         (A)  [right of=l]             {$-2$};
\node[state]         (B) [right of=A] {$-1$};
\node[state]         (C) [right of=B] {$0$};
\node[state]          (D) [right of=C] {$+1$};
\node[state]          (E) [right of=D]  {$+2$};
\node                   (r)     [right of=E]           {$\cdots$};
\path (l)     edge                  [bend right]      node[below]        {\small $1/3$}              (A);
\path (A)     edge                  [bend right]      node[above]        {\small $1/3$}              (l);
\path (A) edge  [bend right] node[below] {\small $1/3$} (B);
\path (A) edge[line width=1.20pt,loop above] node {\small $1/3$} (A);
\path (B) edge[line width=1.20pt,loop above] node {\small $1/3$} (B);
\path (C) edge[line width=1.20pt,loop above] node {\small $1/3$} (C);
\path (D) edge[line width=1.20pt,loop above] node {\small $1/3$} (D);
\path (E) edge[line width=1.20pt,loop above] node {\small $1/3$} (E);
\path (B) edge  [bend right] node[above] {\small $1/3$} (A);
\path (B) edge  [bend right] node[below] {\small $1/3$} (C);
\path (C) edge   [bend right] node[above] {\small $1/3$} (B);
\path (C) edge   [bend right] node[below] {\small $1/3$} (D);
\path (D) edge   [bend right] node[above] {\small $1/3$} (C);
\path (D) edge  [bend right] node[below] {\small $1/3$} (E);
\path (E) edge  [bend right] node[above] {\small $1/3$} (D);
\path (E) edge  [bend right] node[below] {\small $1/3$} (r);
\path (r) edge  [bend right] node[above] {\small $1/3$} (E);
\end{tikzpicture}
\caption{A lazy random walk on the line} \label{LRWonLine}
\end{center}
\end{figure}
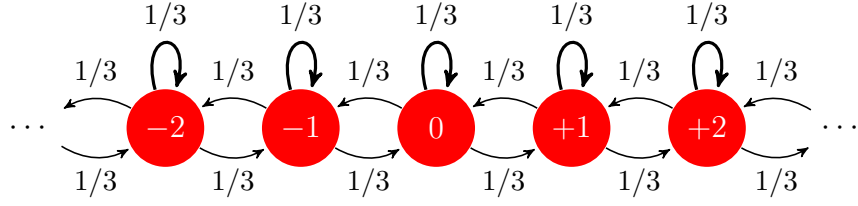

MCBQW can be extended to infinite lattices. As far as we are aware, there is only one publication in the literature concerning MCBQW on an infinite lattice \cite{S13}, where the relationship between the asymptotic average probability distribution of MCBQW and the recurrence of the underlying random walk on the half line is discussed. No publication has ever treated weak limits, the fundamental statistical property for MCBQW on an infinite lattice. In this work we confine our attention to one-dimensional lattice (a line). The underlying Markov chain is a lazy random walk (see Figure \ref{LRWonLine}). 

Let $P$ denote the governing probability operator for the random walk. The transition rules of $P$ are as follows:
\begin{equation}
P|x\rangle=\frac{1}{3}|x-1\rangle+\frac{1}{3}|x\rangle+\frac{1}{3}|x+1\rangle\, \mathrm{for}\, x\in \mathbb{Z}.
\end{equation}

The above equation can be interpreted in this way: 
a walker jumps on the line. At every time step, if he is at location $x$, then
with probability $\frac{1}{3}$ he goes to location $x-1$, with probability $\frac{1}{3}$ to location $x+1$, and with probability $\frac{1}{3}$ stays at location $x$.

In this scenario, the vector states, the orthogonal projector and the swap operator are given as follows:
$|\psi_x\rangle=\frac{\sqrt{3}}{3}|x\rangle\otimes |x-1\rangle+\frac{\sqrt{3}}{3}|x\rangle\otimes |x\rangle+\frac{\sqrt{3}}{3}|x\rangle\otimes |x+1\rangle$ for $x\in \mathbb{Z}$, $\Pi=\sum_{x=-\infty}^{\infty}|\psi_x\rangle\langle \psi_x|$, $S=\sum_{x,y}|x\otimes y\rangle\langle y\otimes x|$.
 

The unitary operator $U=S(2\Pi-1)$ for MCBQW on the line can be expressed by the following formula:

$U(|x\rangle \otimes |x-1\rangle)=-\frac{1}{3}|x-1\rangle \otimes |x\rangle+\frac{2}{3}|x\rangle \otimes |x\rangle+\frac{2}{3}|x+1\rangle \otimes |x\rangle$,
$U(|x\rangle \otimes |x\rangle)=\frac{2}{3}|x-1\rangle \otimes |x\rangle-\frac{1}{3}|x\rangle \otimes |x\rangle+\frac{2}{3}|x+1\rangle \otimes |x\rangle$,
$U(|x\rangle \otimes |x+1\rangle)=\frac{2}{3}|x-1\rangle \otimes |x\rangle+\frac{2}{3}|x\rangle \otimes |x\rangle-\frac{1}{3}|x+1\rangle \otimes |x\rangle.$


The ``overall" state space of the system is $\mathcal{H}=\mathrm{span}\{|x\rangle\otimes |y\rangle, x,y \in \mathbb{Z}\}$ in terms of which a general state of the system may be expressed by the formula:

\[|\psi\rangle=\sum_{x\in\mathbb{Z}} \sum_{y\in \mathbb{Z}}\psi(x,y)|x\rangle\otimes|y\rangle.\]

Given $|\psi_0\rangle \in \mathcal{H}$, where $|||\psi_{0}\rangle||=1$, the expression $|\psi_t\rangle= U^t |\psi_0\rangle$ is the state of the MCBQW at time $t$.
Let $|\psi_t \rangle=\sum_{x \in \mathbb{Z}}[\psi_{t}(x, x-1)|x\rangle\otimes |x-1\rangle+\psi_{t}(x, x)|x\rangle\otimes |x\rangle+\psi_{t}(x, x+1)|x\rangle\otimes |x+1\rangle]$ be the wave function for the MCBQW at time $t$. Then the probability $p_t(x)$ of finding the walker at the position $x$ at time $t$ is given by the standard formula 
$$p_t(x)=|\psi_t (x, x-1)|^2+|\psi_t (x, x)|^2+|\psi_t (x, x+1)|^2,$$
where $|\cdot|$ indicates the modulus of a complex number.

Let $\Psi_{t}(x)\equiv [\psi_t (x, x-1),\psi_t (x, x),\psi_t (x, x+1)]^T$ represent the amplitude of the wave function of the MCBQW at position $x$ and time $t$.

The spatial Fourier transform of $\Psi_{t}(x)$ is defined by
$$ \widehat{\Psi_t}(k)=\sum_{x\in \mathbb{Z}}\Psi_{t}(x)e^{ikx}.$$

Thus, given the initial state $\widehat{\Psi_{0}}(k)$, the Fourier dual of the wave function of the MCBQW system is expressed by
\begin{equation}
\widehat{\Psi_t}(k)=U(k)^t\widehat{\Psi_{0}}(k), \label{eqnwvec}
\end{equation}
where the total evolution operator $U(k)$ is given by 

\begin{equation}
U(k)=\left[\begin{array}{ccc}
0& 0& e^{ik}\\
0&1&0\\
e^{-ik}& 0&0
\end{array}\right]\left[\begin{array}{ccc}
-1/3&2/3&2/3\\
2/3&-1/3&2/3\\
2/3& 2/3&-1/3
\end{array}\right]\label{encodeU}
\end{equation}

The mechanism given in Eq. (\ref{encodeU}) is similar to the one employed by Grimmett {\it et al.}  \cite{GJS04}. Simply speaking, the term $e^{ik}$ encodes the action of a waker jumping from location $x$ to location $x+1$ after one time step, the term $e^{-ik}$ encodes the action of a waker jumping from location $x$ to location $x-1$ after one time step, and the term $e^{0\cdot ik}=1$ encodes the action of a waker staying at location $x$ after one time step.

The above formulation for the unitary operator $U$ lends us a tool to tackle the weak limit of the QW. In what follows, we shall investigate weak limits and limiting distributions of MCBQW on the line.

The three eigenvalues of $U(k)$ are $\lambda_1=-1$, $\lambda_2=\frac{1}{3}+\frac{2}{3}\cos k+\frac{2}{3}i\sqrt{2-\cos^2k-\cos k}$, and $\lambda_3=\frac{1}{3}+\frac{2}{3}\cos k-\frac{2}{3}i\sqrt{2-\cos^2k-\cos k}$. Therefore we have 
\begin{equation}
\overline{\lambda_1}D\lambda_1=0, \, \overline{\lambda_2}D\lambda_2=-\overline{\lambda_3}D\lambda_3=\frac{\sin k}{\sqrt{2-\cos^2k-\cos k}}\label{keyoperator}
\end{equation}
Here, $D=-\mathrm {i} \mathrm{d}/\mathrm{d} k$ denote the position operator in $k$-space.
The corresponding unit eigenvectors are given below:

\begin{equation}
v_1=\frac{1}{\sqrt{4+2\cos k}}\left[\begin{array}{c}
e^{ik}\\
-1-e^{ik}\\
1
\end{array}\right],v_2=\frac{1}{\sqrt{c_2}}\left[\begin{array}{c}
e^{ik}[-3-2\cos k-\frac{2\sqrt{2-\cos^2k-\cos k}\sin k}{1-\cos k}]\\
(-\sin k-\sqrt{2-\cos^2k-\cos k})(\frac{\sin k}{1-\cos k}+i)\\
1
\end{array}\right],\\  \label{vec.-eigen.1}
\end{equation}

\begin{equation}
v_3=\frac{1}{\sqrt{c_3}}\left[\begin{array}{c}
e^{ik}[-3-2\cos k+\frac{2\sqrt{2-\cos^2k-\cos k}\sin k}{1-\cos k}]\\
(-\sin k+\sqrt{2-\cos^2k-\cos k})(\frac{\sin k}{1-\cos k}+i)\\
1
\end{array}\right].\\ \label{vec.-eigen.2}
\end{equation}

Here, $c_2=12+4\cos 2k+12\cos k-8\sin k\sqrt{2-\cos^2 k-\cos k}+\frac{16\sin^2 k+24\sin k \sqrt{2-\cos^2 k-\cos k}}{1-\cos k}$, and 
$c_3=12+4\cos 2k+12\cos k+8\sin k\sqrt{2-\cos^2 k-\cos k}+\frac{16\sin^2 k-24\sin k \sqrt{2-\cos^2 k-\cos k}}{1-\cos k}$.

According to the methods in \cite{GJS04}, the moments of the position distribution are given as 
\begin{eqnarray}
E(X_t^r)=\int_{0}^{2\pi}\langle \widehat{\Psi_t}(k), D^r \widehat{\Psi_t}(k) \rangle \frac{d k}{2\pi}.
\end{eqnarray}

Using the standard calculations, we arrive at, as $t\rightarrow \infty$,
\begin{eqnarray}
E[(X_t/t)^r]=\int_0^{2\pi}\sum_{j=1}^3(\frac{D\lambda_j(k)}{\lambda_j(k)})^r|\langle v_j(k), \widehat{\Psi_0}(k)\rangle|^2\frac{dk}{2\pi}+O(t^{-1}).
\end{eqnarray}

By the method of moments (see \cite{GJS04} and references therein), we can derive the following limit theorem.

Theorem 3. \, Suppose the MCBQW, induced by the lazy random walk on the line, is launched from the origin in the initial state $|\Psi_{0}\rangle=\alpha|0\rangle \otimes |-1\rangle+\beta|0\rangle \otimes |0\rangle+\gamma|0\rangle \otimes |1\rangle$, where $|\alpha|^2+|\beta|^2+|\gamma|^2=1$. For $y\in [-1,1]$, let $\delta_0(y)$ denote the \textit{point mass at the origin} and let $I_{(a,b)}(y)$ denote the \textit{indicator function} of the real interval $(a,b)$. Then, as $t\rightarrow \infty$, the normalized position distribution $f_{t}(y)$ associated with $\frac{1}{t}X_{t}$ converges, in the sense of a weak limit, to the density function
\vskip -0.5cm
\begin{eqnarray}
f(y)=c\delta_0(y)+\frac{I_{(-\frac{\sqrt{6}}{3}, \frac{\sqrt{6}}{3})}(y)}{2\pi (1-y^2)\sqrt{2-3y^2}}(\sum_{j=0}^2a_jy^{j}). \label{densityfunc}
\end{eqnarray}
\vskip -0.2cm
In the above formula, the coefficients $c$, $a_{0}$, $a_{1}$ and $a_{2}$ are given by  
\vskip -0.5cm
\begin{equation}
\left\{
\begin{array}{l}
c=\frac{\sqrt{3}}{6}+\frac{\sqrt{3}-3}{3}\mathrm{Re}(\alpha\overline{\beta})+\frac{3-2\sqrt{3}}{3}\mathrm{Re}(\alpha\overline{\gamma})+\frac{\sqrt{3}-3}{3}\mathrm{Re}(\beta\overline{\gamma})+\frac{2-\sqrt{3}}{2}|\beta|^2 \nonumber\\
a_0=1+|\beta|^2+2\mathrm{Re}(\alpha\overline{\gamma})\nonumber\\
a_1=2|\alpha|^2-2|\gamma|^2+2\mathrm{Re}(\alpha\overline{\beta})-2\mathrm{Re}(\beta\overline{\gamma})\nonumber\\
a_2=1-3|\beta|^2+2\mathrm{Re}(\alpha\overline{\beta})-4\mathrm{Re}(\alpha\overline{\gamma})+2\mathrm{Re}(\beta\overline{\gamma})\nonumber
\end{array}\right.
\end{equation}
Where $\mathrm{Re}(z)$ is the real part of a complex number $z$.
\vskip0.1in

It is noteworthy that a similar type of quantum walk on the line driven by the Grover coin has been studied, and its weak limit was obtained \cite{IKS2005, FB2014}. We would like to point out that MCBQW appears to spread faster than the coin-driven quantum walk although the weak limits of these two types of quantum walks are similar. The aforesaid claim is based on the following simple observation: The indicator function in a formula of density function shows the interval over which the quantum walks prevails. The indicator function of MCBQW discussed in this paper has a wider interval than its counterpart-the quantum walks on the line driven by the Grover coin.

Another attention we would like to draw to MBQW is that this type of quantum walks may have a phenomenon called {\it localization} due to the degeneration of eigenvalues of the time evolution operator ($\lambda_1=-1$) \cite{IKK2004}. We stress that the degeneration of eigenvalues is only the necessary condition for {\it localization}, which in fact also depends upon the initial state of MCBQW. For instance, we consider the case when $\alpha=\beta=\gamma=\frac{\sqrt{3}}{3}$. A direct calculation shows that the density function in Theorem 3 becomes
\begin{equation}
f(y)=\frac{I_{(-\frac{\sqrt{6}}{3}, \frac{\sqrt{6}}{3})}(y)}{\pi (1-y^2)\sqrt{2-3y^2}}.
\end{equation}
This is the case where {\it localization} does not occur as the coefficient $c=0$ in Eq. (\ref{densityfunc}).

Proof of theorem 3.\, We begin with the moments of the position distribution:
\begin{eqnarray}
E[(X_t/t)^r]=\int_0^{2\pi}\sum_j(\frac{D\lambda_j(k)}{\lambda_j(k)})^r|\langle v_j(k), \widehat{\Psi_0}(k)\rangle|^2\frac{dk}{2\pi}+O(t^{-1}).
\end{eqnarray}

By the method of moments (see \cite{GJS04} and references therein), the weak limit of $X_t/t$ exists. Let $Y$ be this weak limit. Then we have 
\begin{eqnarray}
\mathrm{P}(Y \le y)=\int_{h^{-1}(k,j)((-\infty, y])}\sum_{j=1}^3|\langle v_j(k), \widehat{\Psi_0}(k)\rangle|^2\frac{dk}{2\pi}\label{density-fun}
\end{eqnarray}
where $h(k,j)=\overline{\lambda_j}D\lambda_j(k)$ given by Eq. (\ref{keyoperator}). 

According to  Eq. {(\ref{keyoperator}), the probability distribution function in Eq. (\ref{density-fun}) can be written as 
\begin{eqnarray}
\!\!\!\mathrm{P}(Y \le y)
\!\!&=&\!\!H(y)\int_0^{2\pi}|\langle v_1(k), \widehat{\Psi_0}(k)\rangle|^2\frac{dk}{2\pi}\nonumber\\
\!\!&+&\!\!\int_{\arccos\frac{2y^2-1}{1-y^2}}^{2\pi}|\langle v_2(k), \widehat{\Psi_0}(k)\rangle|^2\frac{dk}{2\pi}\nonumber\\
\!\!&+&\!\!\int_0^{2\pi-\arccos\frac{2y^2-1}{1-y^2}}|\langle v_3(k), \widehat{\Psi_0}(k)\rangle|^2\frac{dk}{2\pi}, \, \mathrm{for}\, y\ge 0. \label{distribution}
\end{eqnarray}

\begin{eqnarray}
\!\!\!\mathrm{P}(Y \le y)
\!\!&=&\!\!H(y)\int_0^{2\pi}|\langle v_1(k), \widehat{\Psi_0}(k)\rangle|^2\frac{dk}{2\pi}\nonumber\\
\!\!&+&\!\!\int_{2\pi-\arccos\frac{2y^2-1}{1-y^2}}^{2\pi}|\langle v_2(k), \widehat{\Psi_0}(k)\rangle|^2\frac{dk}{2\pi}\nonumber\\
\!\!&+&\!\!\int_0^{\arccos\frac{2y^2-1}{1-y^2}}|\langle v_3(k), \widehat{\Psi_0}(k)\rangle|^2\frac{dk}{2\pi},  \, \mathrm{for}\, y<0. \label{distribution2}
\end{eqnarray}

Here $H(y)$ is {\em Heaviside function}, which is the cumulative distribution function of $\delta_0(y)$.

After taking derivatives of both sides of Eqs. (\ref{distribution}) and (\ref{distribution2}) with respect to $y$, we obtain the density (in both cases when $y\ge 0$ and $y<0$) as follows
\begin{eqnarray}
\!\!\!f(y)
\!\!&=&\!\!\frac{d\mathrm{P}(Y\le y)}{dy}\nonumber\\
\!\!&=&\!\!\delta_0(y)\int_0^{2\pi}|\langle v_1(k), \widehat{\Psi_0}(k)\rangle|^2\frac{dk}{2\pi}  \label{density2} \nonumber\\
\!\!&+&\!\!\frac{1}{\pi(1-y^2)\sqrt{2-3y^2}}|\langle v_2(k), \widehat{\Psi_0}(k)\rangle|^2_{k=\arccos \frac{2y^2-1}{1-y^2}} \label{density1} \nonumber\\ 
\!\!&+&\!\!\frac{1}{\pi(1-y^2)\sqrt{2-3y^2}}|\langle v_3(k), \widehat{\Psi_0}(k)\rangle|^2_{k=2\pi-\arccos\frac{2y^2-1}{1-y^2}}   \label{density}
\end{eqnarray}

Applying Eqs. (\ref{vec.-eigen.1}) and (\ref{vec.-eigen.2}) to simplify Eq. (\ref{density}), one can obtain the density function $f(y)$ given in Eq. (\ref{densityfunc})

\section{Outlook}

The application of MCBQW to transport for large classes of physical phenomena involving different types of networks has turned out to be successful in recent years such as \cite{PM12, PMCM13}.  Except for Theorems 1 and 2 shown in this article, however, only little is known about the detailed relations between properties of underlying Markov chains and the asymptotic average probability distribution of MCBQW. Therefore, a thorough investigation of the influence of different properties of Markov chains aspects on the dynamics is clearly necessary.

Based on the studies shown above, one may proceed to investigate the things outlined below: 


1. If the Markov chain is irreducible (the directed graph is strongly connected) and reversible, is the asymptotic average probability distribution of MCBQW identical to the stationary probability of MC. Actually, {\em irreducibility} guarantees the existence and uniqueness of the stationary distribution of MC. The condition can be relaxed, we then may restate this conjecture in the following manner: {\em If the Markov chain has a unique stationary probability distribution and is reversible, then the asymptotic average probability distribution of MCBQW is identical to the stationary distribution of MC}.

2. If the Markov chain is irreducible and is not reversible, then is the asymptotic average probability distribution of MCBQW uniform?

\section*{Acknowledgments}

CL thanks US Army Research Laboratory where part of this work was performed, for its hospitality and financial support.
SEVA gratefully acknowledges the financial support of Tecnologico de Monterrey, Escuela de Ingenieria y Ciencias and CONACyT (Fronteras de la Ciencia - project 1007 and SNI member number 41594).


\end{document}